\documentclass[reprint, superscriptaddress, amssymb, amsmath, prb,cha]{revtex4-1}
\usepackage{graphicx}
\usepackage{bm}
\usepackage{soul}

\usepackage[colorlinks=true,linkcolor=blue,urlcolor=blue]{hyperref}
\expandafter\ifx\csname package@font\endcsname\relax\else
 \expandafter\expandafter
 \expandafter\usepackage
 \expandafter\expandafter
 \expandafter{\csname package@font\endcsname}
\fi
\usepackage{color}

\begin{document}

%--------------------------- TITLE --------------------------------------
\title{Defects and Oxygen Impurities in Ferroelectric Wurtzite Al$_{1-x}$Sc$_x$N Alloys}
%----------------------------------------------------------------------------

%----------------------- AUTHORS -------------------------------------
\author{Naseem Ud Din}
\affiliation{Colorado School of Mines, Golden, Colorado 80401, USA}%

\author{Cheng-Wei Lee}
\affiliation{Colorado School of Mines, Golden, Colorado 80401, USA}%

\author{Geoff L. Brennecka}
\affiliation{Colorado School of Mines, Golden, Colorado 80401, USA}%

\author{Prashun Gorai}
\affiliation{Colorado School of Mines, Golden, Colorado 80401, USA}%
\affiliation{National Renewable Energy Laboratory, Golden, CO 80401, USA}
\email{pgorai@mines.edu}

%-------------------------- ABSTRACT -------------------------------
\begin{abstract}
III-nitrides and related alloys are widely used for optoelectronics and as acoustic resonators. Ferroelectric wurtzite nitrides are of particular interest because of their potential for direct integration with Si and wide bandgap semiconductors, and unique polarization switching characteristics; such interest has taken off since the first report of ferroelectric Al$_{1-x}$Sc$_x$N alloys. However, the coercive fields needed to switch polarization are on the order of MV/cm, which is 1-2 orders of magnitude larger than oxide perovskite ferroelectrics. Atomic-scale point defects are known to impact the dielectric properties, including breakdown fields and leakage currents, as well as ferroelectric switching. However, very little is known about the native defects and impurities in Al$_{1-x}$Sc$_x$N, and their effect on the dielectric properties. In this study, we use first-principles calculations to determine the formation energetics of native defects and unintentional oxygen incorporation in Al$_{1-x}$Sc$_x$N. We find that nitrogen vacancies are the dominant native defects, and that they introduce multiple mid-gap states that can lead to premature dielectric breakdown in ferroelectrics and carrier recombination in optoelectronics. Growth under N-rich conditions will reduce the concentration of these deep defects. We also investigate unintentional oxygen incorporation on the nitrogen site and find that the substitutional defect is present in high concentrations, which can contribute to increased temperature-activated leakage currents. Our findings provide fundamental understanding of the defect physics in Al$_{1-x}$Sc$_x$N alloys, which is critical for future deployment of ferroelectric devices.
\end{abstract}
\maketitle

%----------
It has been long known that wurtzite nitrides, including AlN and GaN, have large spontaneous polarization $>$ 100 $\mu$C/cm$^2$, but the electric fields needed to reverse the polarization (for ferroelectricity) are larger than the dielectric breakdown fields.\cite{Dreyer_PRX_2016, Moriwake_APLMater_2020} There has been a resurgence of interest in wurtzite and wurtzite-type ferroelectrics since the unexpected demonstration of robust polarization switching in Al$_{1-x}$Sc$_x$N alloys.\cite{Fichtner_JAP_2019} Polarization switching has now been demonstrated in other solid solutions, including Al$_{1-x}$B$_x$N, Al$_{1-x}$Y$_x$N, Ga$_{1-x}$Sc$_x$N, and Zn$_{1-x}$Mg$_x$O.\cite{hayden_PRM_2021, ferri_2021_JAP, Wang_APL_2021, Wang_APL_2023} New binary and multinary compounds have also been proposed for wurtzite-type ferroelectrics.\cite{Moriwake_APLMater_2020, Dai_SA_2023,lee2023}

There are both scientific and technological reasons to be excited about this new class of ferroelectric materials -- from the fundamentally new physical mechanisms driving polarization switching\cite{Mulaosmanovic_nanoscale_2018, Yazawa_MH_2023, calderon2023} to the possibility of direct integration with commercial microelectronics enabling in-memory computing, high-density data storage, and electro-optics.\cite{Mikolajick_JAP_2021, Kim_nature_nanotech_2023, Mikolajick_JAP_2021} Yet, there are several challenges facing the deployment of wurtzite-type ferroelectric devices, among which lowering of the switching barrier (i.e., coercive field) while remaining highly insulating, is the most pressing. Empirically, coercive field reduction has so far been achieved via strain engineering,\cite{Fichtner_JAP_2019, Yazawa_APL_2021} increased operating temperature,\cite{Zhu_APL_2021} and with higher alloying substitutions. However, significant experimental data and phenomenological modeling both suggest that these approaches have limited impact.\cite{Yazawa_APL_2022} A more rational approach to lowering coercive fields while also maintaining high breakdown fields will require an understanding of the atomic-scale mechanisms.
 
There is ample evidence on the influence of defects on ferroelectric behavior in oxide perovskites,\cite{gao2011revealing} and emerging materials such as HfO$_2$.\cite{chouprik2021defects} However, the defect makeup and their effects on the ferroelectric properties of wurtzite-type compounds and solid solutions remain unclear. We know from fundamental physics that defects, including atomic-scale point defects to extended defects such as stacking faults and dislocations, are likely to affect ferroelectric properties. Charged shallow defects create free electronic carriers (electrons, holes) that contribute to leakage current; in contrast, deep defects localize the electronic carriers and minimize the increase in leakage current but may lead to premature dielectric breakdown. Point defects are also known to locally modify the polarization switching and affect domain wall motion by pinning in oxide ferroelectrics,\cite{otonicar2022} but their role(s) on wurtzite polarization reversal are unknown. 

%-----------------
\begin{figure*}[!t]
\centering
\includegraphics[width=\textwidth]{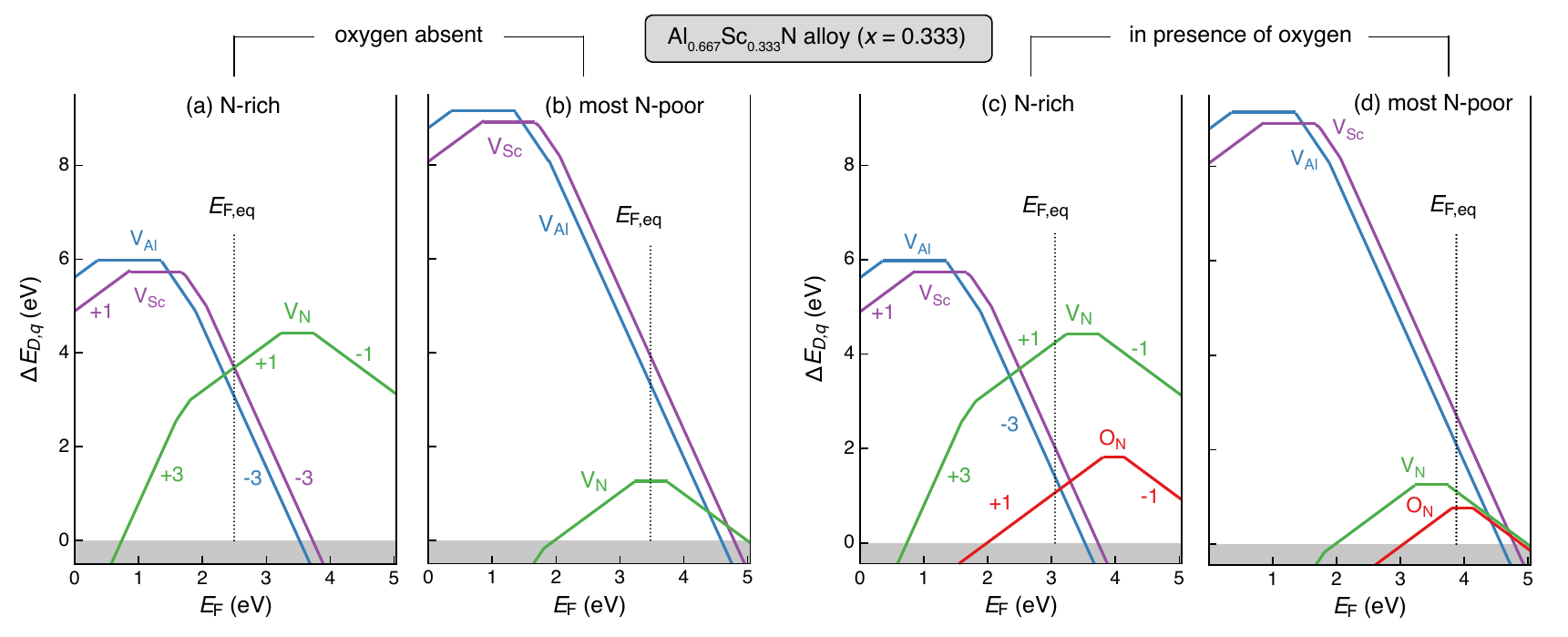}
\caption{\label{fig:defectdiagrams} Defect formation energy ($\Delta E_{D,q}$) as a function of the Fermi energy ($E_\mathrm{F}$) in Al$_{0.667}$Sc$_{0.333}$N alloy when oxygen is absent (a, b) and present (c, d). The equilibrium Fermi energy ($E_{\mathrm{F,eq}}$) is calculated at $T$ = 1000 K. Defect energetics are calculated under N-rich (a, c) and most N-poor growth conditions (b, d) corresponding to the chemical potentials at vertices V4 and V1 of the phase stability region in Table S5, respectively.}
\end{figure*}
%---------------

Our current understanding of the defect physics in wurtzite-type materials is mostly based on binary wurtzite compounds, e.g., GaN and AlN.\cite{Mattila_PRB_1996, yan2014, Gordon_PRB_2014, Osetsky_PRMater_2022} For isostructural alloys between wurtzite compounds, linear interpolation between parent compounds is generally assumed.\cite{Gordon_PRB_2014} However, for heterostructural alloys such as Al$_{1-x}$Sc$_x$N (wurtzite AlN and rocksalt ScN), the validity of such interpolation is unclear. A experimental study proposed nitrogen vacancies as a possible origin of leakage current in Al$_{1-x}$Sc$_{x}$N thin films.\cite{kataoka_JJAP_2021} A computational study probed the formation thermodynamics of nitrogen vacancy ($V_\mathrm{N}$), which qualitatively supports its contribution to the leakage current in Al$_{1-x}$Sc$_x$N.\cite{wang2022} However, the computational study considered only single Al$_{1-x}$Sc$_{x}$N composition ($x$ = 0.25) and did not consider cation vacancies, limits on the elemental chemical potentials set by phase stability, or quantify the defect or electronic carrier concentrations. These details are critical for quantitative predictions.

Additionally, while the majority of studies do not explicitly address oxygen incorporation, it is an experimental reality that all samples have a non-zero amount of oxygen incorporation. Oxygen substituting on the nitrogen site (O$_\mathrm{N}$) is a known donor defect in III-nitrides, and a DX center in some cases e.g., in AlN and related alloys.\cite{Gordon_PRB_2014, yan2014, lyons2017} Therefore, quantifying O$_\mathrm{N}$ and understanding its behavior is critical since shallow donors generate free electrons and contribute to increased leakage currents while deep defects can cause premature dielectric breakdown. Here, we use first-principles defect calculations to determine the formation thermodynamics of native defects and unintentional oxygen impurity incorporation in Al$_{1-x}$Sc$_{x}$N alloys ($x =$ 0 -- 0.333). Specifically, we want to answer: (1) Which defects are present in high concentration and under what growth conditions? (2) Are deep defects present and what are their energetic location? (3) What is the level of unintentional oxygen incorporation -- common in many nitrides and nitride-based alloys? (4) What are the associated electronic carrier concentrations?  

To address these questions, we used our methodology for modeling defects in alloys.\cite{qu_MH_2022, nd_chem_2023} Total energy calculations and structural relaxations were performed with density functional theory (DFT), as implemented in Vienna Ab Initio Software Package (VASP). Details of the computational methodology are provided in the SI. We calculated the formation energetics of native defects and oxygen incorporation in Al$_{1-x}$Sc$_{x}$N with varying Sc content in the range $x =$ 0 -- 0.33, where the wurtzite phase is stable and includes the range of $x$ where ferroelectric switching is experimentally observed. Specifically, we examined the defect properties at $x$ = 0.042, 0.125, 0.250, 0.333, all in the wurtzite phase.    

%--------- Results and Discussion  ----------------

The computed effective formation energy ($\Delta E_{D,q}$) of native defects and substitutional oxygen (O$_\mathrm{N}$) are shown in Figure \ref{fig:defectdiagrams} for $x$ = 0.333. Here, we focus on Al$_{0.667}$Sc$_{0.333}$N, i.e., $x$ = 0.333, only because the qualitative trends across different compositions are similar and compositions with $x > 0.3$ are most relevant for ferroelectric studies. $\Delta E_{D,q}$ is computed for Al vacancy ($V_\mathrm{Al}$), Sc vacancy ($V_\mathrm{Sc}$) and nitrogen vacancy ($V_\mathrm{N}$) defects. We did not consider metal/N antisite defects and interstitials because of their high formation energy in AlN.\cite{Osetsky_PRMater_2022} The equilibrium Fermi energy ($E_\mathrm{F,eq}$) in each case is calculated self-consistently by solving for charge neutrality at 1000 K. Calculated $\Delta E_{D,q}$ at $x$ = 0.042, 0.125, and 0.250 are presented in Figures S4 -- S6. 

\begin{figure*}[!t]
\centering
\includegraphics[width=\textwidth]{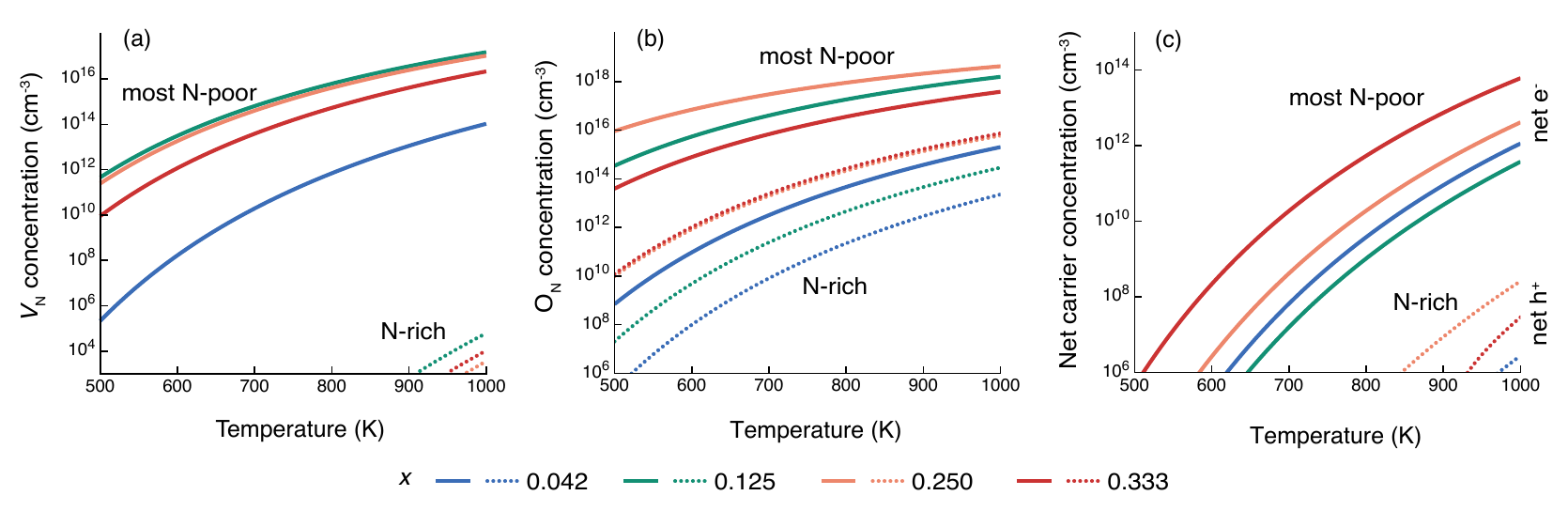}
\caption{\label{fig:concentration} Net carrier concentration in Al$_{1-x}$Sc$_x$N for $x$ = 0.042, 0.125, 0.250, and 0.333 under N-rich and most N-poor growth conditions. (a) Nitrogen vacancy ($V_\mathrm{N}$) concentrations under most N-poor growth conditions (solid lines). Under N-rich conditions (dotted lines), V$_\mathrm{N}$ concentrations are lower $<$ 10$^6$ cm$^{-3}$. (b) Substitutional O$_\mathrm{N}$ concentrations under most N-poor (solid) and N-rich (dotted) growth conditions. (c) Net carrier concentration is $\vert{n_{e} - n_{h}\vert}$, where $n_{e}$ and $n_{h}$ are electron and hole concentrations, respectively. Under most N-poor conditions, $n_e > n_h$ while $n_h > n_e$ under N-rich conditions.}
\end{figure*}

Under N-rich growth conditions (Figure \ref{fig:defectdiagrams}a), $V_\mathrm{Al}$, $V_\mathrm{Sc}$, and $V_\mathrm{N}$ form in low and comparable concentrations due to their very high formation energies; $V_\mathrm{N}$ concentration is $\sim$10$^{4}$ cm$^{-3}$ at 1000 K (Figure \ref{fig:concentration}a). In contrast, we find that $V_\mathrm{N}$ is the dominant defect under the most N-poor growth conditions (Figure \ref{fig:defectdiagrams}a), with equilibrium concentration exceeding 10$^{16}$ cm$^{-3}$. Both metal and anion vacancies exhibit amphoteric behavior, i.e., capable of acting as donor \textit{and} acceptor depending on $E_\mathrm{F}$. High concentrations of $V_\mathrm{N}$ and its amphoteric nature in AlN, GaN and other III-nitrides are well documented.\cite{Lyons_JAP_2021,yan2014,nd_chem_2023} Therefore, it is unsurprising to find that $V_\mathrm{N}$ is the dominant defect, especially under N-poor growth conditions. Our calculations suggest that growth under more N-rich conditions will result in less defective Al$_{1-x}$Sc$_x$N, which is generally desired to sustain larger dielectric breakdown fields, reduced leakage currents, and potentially more robust polarization switching.

%-------------------------
%------------------------

Thin films of III-nitrides and related alloys commonly incorporate oxygen as an impurity.\cite{gungor2022, casamento2020, moram2008} The unintentional oxygen incorporation during growth or ambient exposure has adverse effects on the structural and optoelectronic properties of these materials.\cite{moram2008,yan2014} Specifically, oxygen substituting on the nitrogen site leads to the creation of donor defect states that have been associated with temperature-activated leakage currents in AlN.\cite{schneider2015} Given the possibility of substantial oxygen incorporation in Al$_{1-x}$Sc$_{x}$N alloys, it is crucial to quantify the level of oxygen incorporation and determine if they introduce deep defect states. We calculated $\Delta E_{D,q}$ of oxygen substitution on the nitrogen site (O$_\mathrm{N}$) at each of the four compositions discussed above. Figure \ref{fig:defectdiagrams}(c) and \ref{fig:defectdiagrams}(d) show the defect energetics of O$_\mathrm{N}$ in Al$_{0.667}$Sc$_{0.333}$N, while the defect energetics for $x$ = 0.042, 0.125, and 0.250 are shown in Figures S4, S5, and S6, respectively. Unlike $V_\mathrm{N}$, O$_{\mathrm{N}}$ is the dominant defect (high concentration) under both N-rich and most N-poor growth conditions, suggesting high levels of unintentional O incorporation unless oxygen exposure during growth is carefully eliminated. Similar trends are observed at other compositions (Figures S4 -- S6).

\begin{figure}[!b]
\centering
\includegraphics[width=\linewidth]{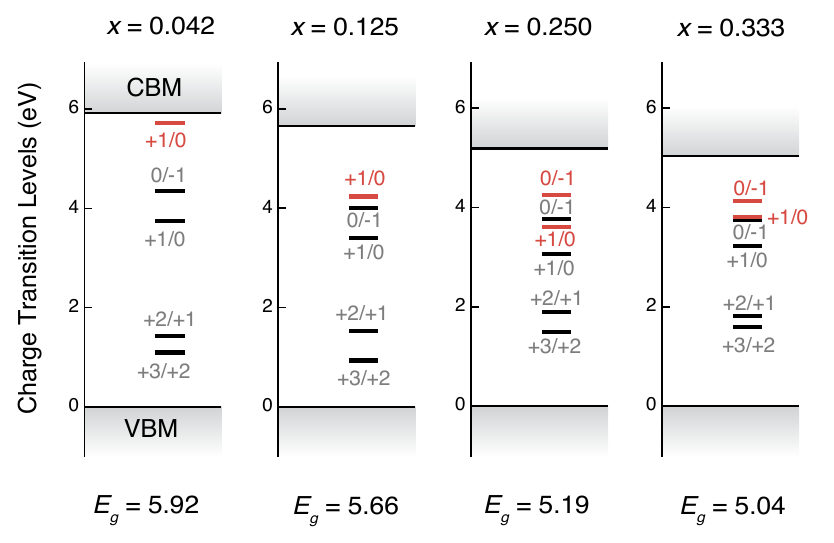}
\caption{\label{fig:ctl} Charge transition levels (CTLs) of nitrogen vacancy (black horizontal bars) and substitutional O$_\mathrm{N}$ (red horizontal bars) in  Al$_{1-x}$Sc$_{x}$N alloys. CTLs are referenced to the valence band maximum (VBM). The band gap ($E_g$) monotonically decreases with increasing Sc content ($x$).}
\end{figure}

We find that both $V_\mathrm{N}$ and O$_\mathrm{N}$ introduce multiple deep mid-gap states in Al$_{1-x}$Sc$_{x}$N. The charge transition levels (CTLs) of $V_\mathrm{N}$ are shown in Figure \ref{fig:ctl} at various alloy compositions. The CTLs are referenced to the valence band maxima (VBM). The band gap is calculated using the GW method (see SI for details), which shows a monotonic decrease with increasing $x$ since ScN has a significantly smaller band gap ($\sim$ 1 eV)\cite{saha2010} relative to AlN (calculated 6.1 eV). O$_\mathrm{N}$ is a known DX center in AlN.\cite{Gordon_PRB_2014} A DX center is a deep defect that forms when a donor defect like O$_\mathrm{N}^{+1}$ captures two electrons, and undergoes a large local structural distortion to stabilize an acceptor state like O$_\mathrm{N}^{-1}$. As a result, a key feature of a DX center is displacement of the O atom from the ideal N site. DX centers are commonly observed in zincblende and wurtzite compounds.\cite{Li_JMCC_2019} For Al$_{1-x}$Sc$_{x}$N alloys, we find O$_\mathrm{N}^{-1}$ has the characteristics of a DX center (e.g., see structures in Figure S7 for $x$ = 0.333). Both deep defects are undesirable for optoelectronics because they act as carrier recombination centers. In ferroelectrics, deep defects can mediate premature dielectric breakdown before polarization switching but can be beneficial for electronic carrier trapping that suppresses leakage currents.

%-------------------
%-------------

The breakdown field of an insulator critically depends on the band gap of the material. Considering the importance of the breakdown field for polarization switching, we semi-quantitatively estimate the effect of Sc composition ($x$) on the intrinsic breakdown field ($E_b$, in MV/m) of Al$_{1-x}$Sc$_x$N. We used a phenomenological model\cite{Kim_CM_2016} to estimate $E_b$ as,
%---------
\begin{equation}\label{eqn:Ebd}
    E_b = 24.442 ~ exp\left( 0.315 \sqrt{E_g \omega_{max}}\right)
\end{equation}
%---------
which depends on the electronic band gap ($E_g$, in eV) and the maximum optical phonon frequency ($\omega_{max}$, in THz) at the $\Gamma$ point of the Brillouin zone. $E_g$ of the alloys calculated with the GW method (Figure \ref{fig:ctl}) monotonically decreases with $x$. Direct calculation of $\omega_{max}$ for alloys is computationally expensive, but we can draw trends based on $\omega_{max}$ of pure wurtzite-AlN ($\sim$26.98 THz) and rocksalt-ScN ($\sim$19.49--20.99 THz). We expect a decrease in $\omega_{max}$ with increasing $x$, consistent with the increased softening of Al$_{1-x}$Sc$_x$N. With pure AlN ($E_g$ = 6.11 eV) as the reference and no change in $\omega_{max}$, $E_b$ will decrease by 4\%, 9\%, 18\%, and 20\% at $x$ = 0.042, 0.125, 0.250, and 0.333, respectively. The decrease in $\omega_{max}$ with $x$ will further exacerbate the decrease in $E_b$. 

Charged defects introduce electronic carriers (free or bound) that contribute to leakage currents. We estimate the net carrier concentrations self-consistently by solving for charge neutrality. The net carrier concentration computed as $\vert{n_{h} - n_{e}\vert}$ is shown in Figure \ref{fig:concentration} (c), corresponding to the defect energetics in Figures \ref{fig:defectdiagrams}(a) and (b) where no oxygen incorporation is taken into account. Here, $n_{e}$ and $n_{h}$ are electron and hole concentrations, respectively. The qualitative trends are similar with and without oxygen incorporation. The net carrier concentration is $e^{-}$ for $n_{e} > n_{h}$ and $h^{+}$ for $n_{e} < n_{h}$. We find that, regardless of the composition, the net carrier concentration decreases under N-rich conditions. This decrease in net carrier concentration suggests that growth under N-rich conditions can effectively minimize leakage currents. Even though the net carrier concentration is predicted to be large ($>$10$^{12}$ cm$^{-3}$ for $x \geq$ 0.250) under N-poor conditions, the presence of very deep defect states (Figure \ref{fig:ctl}) means the carriers will be bound and not available as free conduction carriers. Such bound carriers do not contribute to leakage current, and can only be temperature activated, as is observed in pure AlN.\cite{schneider2015}

Future extensions of our work may consider hydrogen defects, e.g., hydrogenated anion vacancies in AlN,\cite{szucs2003} and cation complexes, e.g., Al vacancy complex with oxygen impurity in AlN.\cite{yan2014} The search for these defect complexes is computationally intensive even in ordered compounds and challenging because there is no prescription to systematically identify them.

In summary, our defect calculations reveal that nitrogen vacancies are the dominant defects in Al$_{1-x}$Sc$_x$N alloys, which can be minimized by growth under N-rich conditions. Both metal and anion vacancies introduce multiple deep defect states; it is critical to minimize their concentration by growth under relatively N-rich conditions to reduce carrier recombination for optoelectronics and premature dielectric breakdown in power electronics and ferroelectrics. N-rich growth also reduces net carrier concentrations -- beneficial for reducing leakage currents. High levels of unintentional oxygen incorporation are expected, with substitutional O$_\mathrm{N}$ present in even higher concentrations than $V_\mathrm{N}$ and leading to even higher leakage currents. Systematic elimination of oxygen during growth is, therefore, critical for ferroelectric nitrides.\\ 

\noindent
\textbf{Acknowledgements}
Support for this work was provided by the National Science Foundation under Grant No. DMR-2119281. The work was also partially supported by the Department of Energy Basic Energy Sciences (BES), with additional support from Advanced Scientific Computing Research (ASCR), under program ERW6548. The research was performed using computational resources sponsored by the Department of Energy's Office of Energy Efficiency and Renewable Energy and located at the NREL. The views expressed in the article do not necessarily represent the views of the DOE or the U.S. Government.\\

\noindent
\textbf{Data Availability}
The data that supports the findings of this study are available within the article and its supplementary material.

\section*{Author Declarations}
\noindent
\textbf{Conflict of Interest}

\noindent
The authors have no conflicts to disclose.\\

\noindent
\textbf{Author Contributions}

\noindent
\textbf{Naseem Ud Din}: Investigation, Data Curation, Writing (Original Draft). \textbf{Cheng-Wei Lee}: Investigation, Data Curation, Writing (Editing). \textbf{Geoff L. Brennecka}: Writing (Editing), Project Administration. \textbf{Prashun Gorai}: Conceptualization, Investigation, Data Curation, Writing (Editing), Supervision, Project Administration.

%-------------------------------------------------------------
\bibliography{biblio}
%--------------------------------------------------------------------------

\end{document}

%---------------------------
\renewcommand{\thetable}{S\arabic{table}}
\renewcommand{\thefigure}{S\arabic{figure}}

\clearpage

\section{Computational Methods}

\subsection{Alloy Modeling and Structure Relaxation}\label{sec:alloymodel}
We modeled Al$_{1-x}$Sc$_{x}$N alloys with special quasirandom structures (SQS).\cite{zunger_PRL_1990} We used the \textit{mcsqs} function implemented in the Alloy Theoretic Automated Toolkit (ATAT) to generate the SQS structures.\cite{van_Calphad_2009} We created 48-atom SQS supercells, with pair and triplet clusters considered within a range of up to 4.5~\AA~ and 3.2~\AA, respectively. We fully relaxed the SQS supercells (atomic positions, cell volume, and cell shape) using plane wave density functional theory (DFT), as implemented in Vienna Ab-initio Simulation Package (VASP),\cite{kresse_PRB_1996} using the generalized gradient approximation (GGA) of Perdew-Burke-Ernzerhof (PBE) as the exchange correlation functional. The plane waves were expanded with an energy cutoff of 340 eV. The following PAW potentials\cite{blochl_PRB_1994} were used to treat the core electrons: Al 04Jan2001, Sc\textunderscore sv 07Sep2000, N\textunderscore s 07Sep2000, and O\textunderscore s 07Sep2000. Specifically, we used the GGA(+U) approach with on-site Hubbard $U$ = 3.0 eV correction applied to the Sc $d$ orbitals.\cite{Dudarev_PRB_1998} The SQS structures were relaxed until the residual forces on each atom were below 0.01 eV/\AA.

\subsection{Defect Calculations}\label{sec:defects_calc}

We calculated the defect formation energetics in Al$_{1-x}$Sc$_{x}$N alloys by adapting the standard supercell approach for ordered compounds. This approach is based on creating a supercell of the alloy SQS and calculating the \textit{effective} defect formation energy. We have employed this approach to understand the defect physics of Ba$_{2(1-x)}$Sr$_{2x}$CdP$_2$ Zintl phase alloys and Al$_{1-x}$Gd$_{x}$N alloys. \cite{qu_2022_MH, nd_chem_2023} 

For each alloy composition, i.e., $x$ in Al$_{1-x}$Sc$_{x}$N, we constructed 192-atom supercells of the 48-atom SQS (discussed above). SQS cells have \textit{P1} symmetry, and therefore, each atom position is a unique Wyckoff site. The Wyckoff sites represent different local environments in the alloy and as such, defect formation energy ($\Delta E_{D,q}$) is expected to vary with each Wyckoff site. To build the statistics of defect formation energy in alloys, defects ($D$) at each Wyckoff site must be considered individually, which results in an extremely large number of calculations when each defect is calculated in 5-7 different charge states ($q$). To reduce the computational expense, in our approach, we first calculate $\Delta E_{D,q}$ of Al vacancy ($V_\mathrm{Al}$), Sc vacancy ($V_\mathrm{Sc}$), and N vacancy ($V_\mathrm{N}$) at all Wyckoff sites in a \textit{single} charge state, which is the most favorable assuming nominal oxidation states of each element. Specifically, we calculated V$_\mathrm{Al}$ in $q = -3$, V$_\mathrm{Sc}$ in $q = -3$, and V$_\mathrm{N}$ in $q = +3$. From these calculations, we identified the Wyckoff sites associated with the maximum and minimum $\Delta E_{D,q}$ for each defect. We then calculated $\Delta E_{D,q}$ only at these two sites for each defect but in all plausible charge states, i.e., V$_\mathrm{Al}$ and V$_\mathrm{Sc}$ in $q$ = -3, -2, -1, 0, 1 and  V$_\mathrm{N}$ in $q$ = -1, 0, 1, 2, 3. Ultimately, we calculated the effective defect formation energy ($\Delta E_{D,q}^{eff}$) for each defect type from the minimum ($\Delta E_{D,q}^{min}$) and maximum ($\Delta E_{D,q}^{max}$) values, following the averaging proposed by Zhang, et al.,\cite{zhang_2023_nature}
%--------------
\begin{equation}
exp\left({-\frac{\Delta E_{D,q}^{eff}}{k_\mathrm{B}T}}\right) = \frac{1}{2} \left[ exp \left({-\frac{\Delta E_{D,q}^{min}}{k_\mathrm{B}T}}\right) + exp\left({-\frac{\Delta E_{D,q}^{max}}{k_\mathrm{B}T}}\right) \right]
\label{eq:effectiveformation}
\end{equation}
%-------------
where $k_\mathrm{B}$ and T are the Boltzmann constant and temperature, respectively.

At a given Wyckoff site, $\Delta E_{D,q}$ was calculated using the standard supercell approach,
%---------
\begin{equation}
\Delta E_{D,q} = E_{D,q} - E_{\text{bulk}} + \sum_{i} n_{i} \mu_{i} + qE_{\text{F}} + E_{\text{corr}}
\label{eq:defect_eq}
\end{equation}
%---------
where $E_{D,q}$ and $E_{bulk}$ are the total energies of the defect and the bulk supercell, respectively. $n_{i}$ is the number of atoms of element $i$ added to ($n_{i} < 0$) or removed from ($n_{i} > 0$), the bulk supercell, to create the defect supercell. $\mu_{i}$ is the chemical potential of element $i$. More details on the calculation of $\mu_i$ are discussed in the next section. $E_\mathrm{F}$ is the Fermi energy, which is referenced to the valence band maximum. $E_{\text{corr}}$ term encompasses the finite-size corrections; we followed the correction scheme proposed by Lany and Zunger.\cite{lany_2008_PRB} To address the underestimation of the band gap in DFT, we shifted the DFT-computed band edges based on the electronic band structure calculated with the GW method.\cite{lany_2008_PRB,peng_2013_PRB} The GW calculations were performed on the 48-atom SQS cells for each composition $x$ using a $2\times 2 \times 2$ $k$-point grid. The grid was chosen to accurately capture the $k$-point position of the band edges while balancing the computational cost for the 48-atom cell. The number of unoccupied bands was set to more than 10X the number of occupied bands to obtain converged GW eigenenergies. The calculations were set up and analyzed using the open-source Python package \textit{pylada-defects}.\cite{goyal_2017_CMS} The finite-size corrections were also calculated using \textit{pylada-defects}. 

We calculated the equilibrium Fermi energy $E_{\mathrm{F},eq}(T)$ and the defect $C_D(T)$ and electronic carrier concentrations at a given temperature T. $E_{\mathrm{F},eq}$ was calculated self-consistently using the charge neutrality condition. We used the open-source code \textit{py-sc-fermi} to obtain $E_{\mathrm{F},eq}(T)$.\cite{buckeridge2019equilibrium} $C_D(T)$ was calculated as,
%---------
\begin{equation}
    C_{D}(T) = N_{sites} exp \left[ - \frac{\Delta E_{D,q}^{eff}}{k_\text{B}T}\right]
\end{equation}
%------------
where N$_{sites}$ is the concentration of the structure sites in Al$_{1-x}$Sc$_{x}$N alloys where the defect can form, e.g., N sites for $V_N$.  

%-------------
\subsection{Chemical Potentials and Phase Stability}

The chemical potential of elemental $i$ ($\mu_i$ in Equation \ref{eq:defect_eq}) is expressed relative to the chemical potential in the reference states ($\mu_i^0$). Therefore, $\mu_i = \mu_i^0 + \Delta \mu_i$, where $\Delta \mu_i$ is the deviation from the reference state. Generally, DFT total energy in the bulk elemental form is used as $\mu_i^0$ but is known to inaccurately predict the formation enthalpy and phase stability.\cite{stevanovic2012} Instead, we used $\mu_i^0$ values that were obtained by fitting to a set of experimentally measured formation enthalpy values under standard conditions.\cite{stevanovic2012} The fitted $\mu_i^0$ used in our calculations are listed in Table S1. 

The allowable range of $\Delta \mu_{i}$ is restricted by the thermodynamic phase stability condition. Specifically, the range was calculated via convex hull analyses within the grand canonical ensemble. For alloys, the formation free energy instead of formation enthalpy was used to construct the convex hull. The calculated enthalpy of mixing ($\Delta H_m$) is shown in Figure S1. $\Delta H_m$ was calculated in the wurtzite and rocksalt phases because Al$_{1-x}$Sc$_{x}$N is a heterostructural alloy between wurtzite AlN and rocksalt ScN. We predict a critical composition $x_c = 0.56$ beyond which rocksalt is the ground-state structure of the alloy. Combined with the configurational entropy of ideal mixing, we calculated the free energy of mixing as a function of temperature. We then chose an effective temperature of 3000 K to calculate the formation free energy. $\Delta \mu_i$ ranges were calculated from the convex hull analysis by treating each alloy composition as a line compound.

The calculated $\Delta \mu_{i}$ values for different $x$ at the vertices of the convex hull, which define the region of phase stability in the ternary Al-Sc-N chemical space, are listed in Tables S2-S5. Using composition $x = 0.25$ as an example, Figure S2 shows the ternary phase diagram in the composition space. Similarly, when considering O incorporation, the calculated $\Delta \mu_{i}$ values in the quaternary Al-Sc-N-O chemical space are presented in Tables S6-S9, and Figure S3 shows the quaternary phase diagram for $x = 0.25$. For gaseous elements, we used a rigid-dumbbell ideal gas model to estimate the temperature and partial pressure dependency of the chemical potential. Specifically for oxygen, we used $T$ = 1000 K and oxygen partial pressure ($p_{\mathrm{O}_2}$) of 0.2 atm,\cite{Osorio-Guillen_PRL_2006,nd_chem_2023} 
%---------
\begin{equation}
\mu_\mathrm{O}(p_\mathrm{O_2},T) = \mu_\mathrm{O}^0 + \Delta \mu_\mathrm{O} (T) + \frac{1}{2}k_\mathrm{B}T~\mathrm{ln}\left(p_{\mathrm{O}_2}\right)
\label{eq:mo_temp}
\end{equation}
%---------
where $\mu_\mathrm{O}^0$ is the reference chemical potential for oxygen, and $\Delta \mu_\mathrm{O} (T)$ is defined as,
%---------
\begin{equation}
\Delta \mu_\mathrm{O} (T)  = \frac{1}{2} C_{P}^{0}\left(T -T^0 \right) - \frac{1}{2} T \left[S_{\mathrm{O}_2}^{0} + C_{P}^{0}~\mathrm{ln}\left(\frac{T}{T^0}\right)\right]
\label{eq:delta_temp}
\end{equation}
%---------
where $C_{P}^{0}$ = 3.02$\times$10$^{-4}$ eV/K,  $T^0$ = 298.15 K and $S_{\mathrm{O}_2}^{0}$ = 0.0021 eV/K.\cite{Haynes_2014_crc} For nitrogen, we did not consider the temperature dependency of the chemical potential to approximately represent the more N-rich conditions beyond the thermodynamic limit that can be achieved by non-equilibrium growth using nitrogen plasma.\cite{nd_chem_2023}

\clearpage

\section{Alloy Mixing Enthalpy}

\begin{figure}[!ht]
\centering
\includegraphics[width=0.6\linewidth]{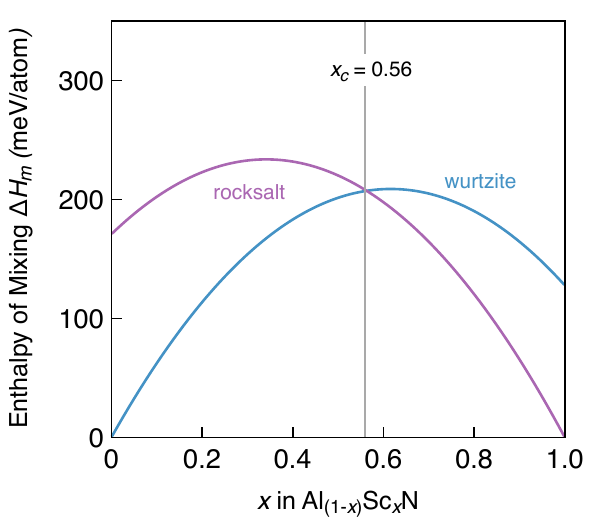}
\caption{\label{fig:enthalpy}
Calculated thermodynamics of heterostructural Al$_{1-x}$Sc$_{x}$N alloys. Mixing enthalpy ($\Delta H_{m}$) curves of Al$_{1-x}$Sc$_{x}$N in wurtzite and rocksalt phases.}
\end{figure}
%-----------------

\section{Elemental Reference Chemical Potentials}
%---------------------------
\begin{table}
\caption{Fitted elemental reference chemical potentials under standard conditions.}
\begin{tabular}{cc}
\hline
& \\
Element & $\mu^0$ (eV) \\
& \\
\hline
& \\
Al & -3.270 \\
Sc & -4.420\\
N & -8.460\\
O & -4.800\\
& \\
\hline
\end{tabular}\\
\label{table:chempot}
\end{table}
%---------------------------

\newpage
\section{Alloy Phase Stability in Ternary Al-Sc-N Chemical Space}

\begin{table}
\caption{Three-phase equilibrium regions of Al$_{0.958}$Sc$_{0.042}$N (Al$_{23}$Sc$_{1}$N$_{24}$) in the ternary Al-Sc-N chemical space. $\Delta \mu_{i}$ ($i$ = Al, Sc, N) are the deviations in the elemental chemical potential from the reference values.}

\begin{tabular}{ccccc}
\hline
&&&&\\
Vertex & $\Delta \mu_{\mathrm{Al}}$ & $\Delta \mu_{\mathrm{N}}$ & $\Delta \mu_{\mathrm{Sc}}$ & Phases in equilibrium with Al$_{23}$Sc$_{1}$N$_{24}$\\
& (eV) & (eV) & (eV)&\\
&&&&\\
\hline
&&&&\\
V1& 0.000 & -3.208 & -1.102&Al, ScN,  Al$_{23}$Sc$_{1}$N$_{24}$ \\
V2& 0.000&-3.172& -1.970& Al, , AlN, Al$_{23}$Sc$_{1}$N$_{24}$\\
V3&-3.172&0.000&-5.142&N$_{2}$, AlN, Al$_{23}$Sc$_{1}$N$_{24}$\\
V4&-3.208&0.000&-4.310&N$_{2}$, ScN, Al$_{23}$Sc$_{1}$N$_{24}$ \\
&&&&\\
\hline
\end{tabular}\\
\label{table:compositionpd04}
\end{table}
%---------------------------

%---------------------------
\begin{table}
\caption{Three-phase equilibrium regions of Al$_{0.875}$Sc$_{0.125}$N (Al$_{21}$Sc$_{3}$N$_{24}$) in the ternary Al-Sc-N chemical space. $\Delta \mu_{i}$ ($i$ = Al, Sc, N) are the deviations in the elemental chemical potential from the reference values.}
\begin{tabular}{ccccc}
\hline
&&&&\\
Corner & $\Delta \mu_{\mathrm{Al}}$ & $\Delta \mu_{\mathrm{N}}$ & $\Delta \mu_{\mathrm{Sc}}$ & Phases in equilibrium with Al$_{21}$Sc$_{3}$N$_{24}$ \\
& (eV) & (eV) & (eV)&\\
&&&&\\
\hline
&&&&\\
V1& 0.000& -3.222 & -1.088&Al,  ScN, Al$_{21}$Sc$_{3}$N$_{24}$ \\
V2& 0.000&-3.172& -1.490&Al,  AlN, Al$_{21}$Sc$_{3}$N$_{24}$ \\
V3&-3.172&0.000&-4.662&N$_{2}$, AlN, Al$_{21}$Sc$_{3}$N$_{24}$\\
V4&-3.222&0.000&-4.310&N$_{2}$,ScN, Al$_{21}$Sc$_{3}$N$_{24}$ \\
&&&&\\
\hline
\end{tabular}\\
\label{table:compositionpd125}
\end{table}
%---------------------------

%---------------------------
\begin{table}
\caption{Three-phase equilibrium regions of Al$_{0.750}$Sc$_{0.250}$N (Al$_{18}$Sc$_{6}$N$_{24}$) in the ternary Al-Sc-N chemical space. $\Delta \mu_{i}$ ($i$ = Al, Sc, N) are the deviations in the elemental chemical potential from the reference values.}
\begin{tabular}{ccccc}
\hline
&&&&\\
Corner & $\Delta \mu_{\mathrm{Al}}$ & $\Delta \mu_{\mathrm{N}}$ & $\Delta \mu_{\mathrm{Sc}}$ & Phases in equilibrium with Al$_{18}$Sc$_{6}$N$_{24}$ \\
& (eV) & (eV) & (eV)&\\
&&&&\\
\hline
&&&&\\
V1&0&-3.201&-1.109&Al, ScN, Al$_{18}$Sc$_{6}$N$_{24}$\\
V2&0&-3.172&-1.226&Al, AlN, Al$_{18}$Sc$_{6}$N$_{24}$ \\
V3& -3.172 & 0&-4.398 & N$_{2}$, AlN, Al$_{18}$Sc$_{6}$N$_{24}$ \\
V4& -3.201&0& -4.310&  N$_{2}$, ScN, Al$_{18}$Sc$_{6}$N$_{24}$ \\
&&&&\\
\hline
\end{tabular}\\
\label{table:compositionpd25}
\end{table}
%----------------------

%---------
\begin{table}
\caption{Three-phase equilibrium regions of Al$_{0.667}$Sc$_{0.333}$N (Al$_{16}$Sc$_{8}$N$_{24}$) in the ternary Al-Sc-N chemical space. $\Delta \mu_{i}$ ($i$ = Al, Sc, N) are the deviations in the elemental chemical potential from the reference values.}
\begin{tabular}{ccccc}
\hline
&&&&\\
Corner & $\Delta \mu_{\mathrm{Al}}$ & $\Delta \mu_{\mathrm{N}}$ & $\Delta \mu_{\mathrm{Sc}}$ & Phases in equilibrium with Al$_{16}$Sc$_{8}$N$_{24}$ \\
& (eV) & (eV) & (eV)&\\
&&&&\\
\hline
&&&&\\
V1&0&-3.174&-1.136&Al, ScN, Al$_{16}$Sc$_{8}$N$_{24}$\\
V2&0&-3.172&-1.142&Al, AlN, Al$_{16}$Sc$_{8}$N$_{24}$ \\
V3& -3.172 & 0&-4.314 & N$_{2}$, AlN, Al$_{16}$Sc$_{8}$N$_{24}$ \\
V4& -3.174&0& -4.310&  N$_{2}$, ScN, Al$_{16}$Sc$_{8}$N$_{24}$ \\
&&&&\\
\hline
\end{tabular}\\
\label{table:compositionpd33}
\end{table}

%-------------

%---------------------------
\begin{figure}[h]
\centering
    \includegraphics[width=0.6\linewidth]{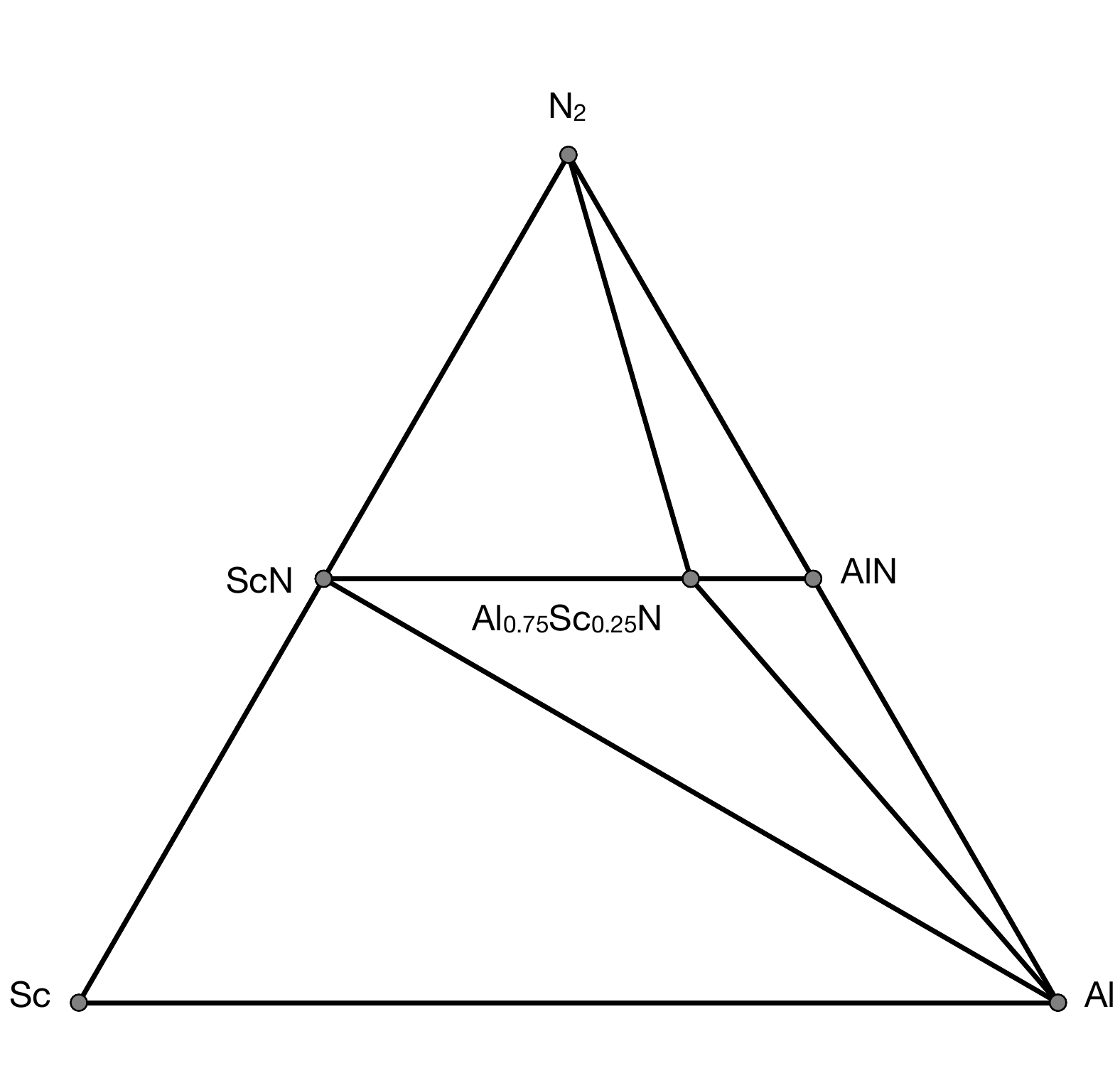}
    \caption{\label{fig:SI_25_phase}
    Calculated Al-Sc-N ternary phase diagram showing Al$_{0.75}$Sc$_{0.25}$N as a line compound.}
\end{figure}

%---------------------------
\clearpage

\section{Alloy Phase Stability in Quaternary Al-Sc-N-O Chemical Space}

\begin{table}
\caption{Four-phase equilibrium regions of Al$_{0.96}$Sc$_{0.04}$N (Al$_{23}$Sc$_{1}$N$_{24}$) in the quaternary Al-Sc-N-O chemical potential space. $\Delta \mu_{X}$ are the deviations in the elemental chemical potential from the reference values.}
\begin{tabular}{cccccc}
\hline
&&&&&\\
Corner & $\Delta \mu_{\mathrm{Al}}$ & $\Delta \mu_{\mathrm{N}}$ & $\Delta \mu_{\mathrm{O}}$ &$\Delta \mu_{\mathrm{Sc}}$ &Phases in equilibrium with Al$_{23}$Sc$_{1}$N$_{24}$ \\
& (eV) & (eV) & (eV)& (eV) &\\
&&&&&\\
\hline
&&&&&\\
V1&0.000&-3.188&-5.520&-1.583&Al, Sc$_{2}$O$_3$,Al$_{2}$O$_3$, Al$_{23}$Sc$_{1}$N$_{24}$ \\
V2&0.000&-3.208&-5.840&-1.102&Al,Sc$_{2}$O$_3$, ScN, Al$_{23}$Sc$_{1}$N$_{24}$ \\
V3&0.000&-3.172&-5.520&-1.970&Al,Al$_{2}$O$_{3}$, AlN, Al$_{23}$Sc$_{1}$N$_{24}$ \\
V4& -3.188 & 0.000 & -3.395&-4.771& N$_{2}$,  Al$_{2}$O$_{3}$, Sc$_{2}$O$_{3}$, Al$_{23}$Sc$_{1}$N$_{24}$ \\
V5& -3.208&0.000& -3.702& -4.310& N$_{2}$, Sc$_{2}$O$_{3}$ , ScN,  Al$_{23}$Sc$_{1}$N$_{24}$ \\
V6&-3.172&0.000&-3.405&-5.142&N$_{2}$, Al$_{2}$O$_3$,  AlN,  Al$_{23}$Sc$_{1}$N$_{24}$\\
&&&&&\\
\hline
\end{tabular}\\
\label{table:compositionpdwithP04}
\end{table}
%-------------------------

%---------------------------
\begin{table}
\caption{Four-phase equilibrium regions of Al$_{0.875}$Sc$_{0.125}$N (Al$_{21}$Sc$_{3}$N$_{24}$) in the quaternary Al-Sc-N-O chemical potential space. $\Delta \mu_{X}$ are the deviations in the elemental chemical potential from the reference values.}
\begin{tabular}{cccccc}
\hline
&&&&&\\
Corner & $\Delta \mu_{\mathrm{Al}}$ & $\Delta \mu_{\mathrm{N}}$ & $\Delta \mu_{\mathrm{O}}$ &$\Delta \mu_{\mathrm{Sc}}$ &Phases in equilibrium with Al$_{21}$Sc$_{3}$N$_{24}$ \\
& (eV) & (eV) & (eV)& (eV) &\\
&&&&&\\
\hline
&&&&&\\
V1&0.000&-3.222&-5.850&-1.088&Al, Sc$_{2}$O$_3$,ScN, Al$_{21}$Sc$_{3}$N$_{24}$ \\
V2&0.000&-3.172&-5.582&-1.490&Al,Sc$_{2}$O$_3$, AlN, Al$_{21}$Sc$_{3}$N$_{24}$ \\
V3&-3.222&0.000&-3.702&-4.310&N$_{2}$,Sc$_{2}$O$_{3}$, ScN, Al$_{21}$Sc$_{3}$N$_{24}$ \\
V4& -3.172&0.000& -3.467 & -4.662& N$_{2}$,  AlN, Sc$_{2}$O$_{3}$, Al$_{21}$Sc$_{3}$N$_{24}$ \\
&&&&&\\
\hline
\end{tabular}\\
\label{table:compositionpdwithP125}
\end{table}

%---------------------------

%---------------------------
\begin{table}
\caption{Four-phase equilibrium regions of Al$_{0.75}$Sc$_{0.25}$N (Al$_{18}$Sc$_{6}$N$_{24}$) in the quaternary Al-Sc-N-O chemical potential space. $\Delta \mu_{X}$ are the deviations in the chemical potential from the reference values.}
\begin{tabular}{cccccc}
\hline
&&&&&\\
Corner & $\Delta \mu_{\mathrm{Al}}$ & $\Delta \mu_{\mathrm{N}}$ & $\Delta \mu_{\mathrm{O}}$ &$\Delta \mu_{\mathrm{Sc}}$ &Phases in equilibrium with Al$_{18}$Sc$_{6}$N$_{24}$ \\
& (eV) & (eV) & (eV)& (eV) &\\
&&&&&\\
\hline
&&&&&\\
V1& 0.000 & -3.201 & -5.836&-1.109& Al, ScN, Sc$_{2}$O$_3$, Al$_{18}$Sc$_{6}$N$_{24}$ \\
V2& 0.000&-3.172& -5.758& -1.226& Al, AlN, Sc$_{2}$O$_3$, Al$_{18}$Sc$_{6}$N$_{24}$ \\
V3&-3.201&0.000&-3.702&-4.310&N$_{2}$, ScN, Sc$_{2}$O$_3$,Al$_{18}$Sc$_{6}$N$_{24}$\\
V4&-3.172&0.000&-3.643&-4.398&N$_{2}$, AlN, Sc$_{2}$O$_3$, Al$_{18}$Sc$_{6}$N$_{24}$ \\
&&&&&\\
\hline
\end{tabular}\\
\label{table:compositionpdwithO25}
\end{table}
%----------
\begin{table}
\caption{Four-phase equilibrium regions of Al$_{0.667}$Sc$_{0.333}$N (Al$_{16}$Sc$_{8}$N$_{24}$) in the quaternary Al-Sc-N-O chemical potential space. $\Delta \mu_{X}$ are the deviations in the chemical potential from the reference values.}
\begin{tabular}{cccccc}
\hline
&&&&&\\
Corner & $\Delta \mu_{\mathrm{Al}}$ & $\Delta \mu_{\mathrm{N}}$ & $\Delta \mu_{\mathrm{O}}$ &$\Delta \mu_{\mathrm{Sc}}$ &Phases in equilibrium with Al$_{16}$Sc$_{8}$N$_{24}$ \\
& (eV) & (eV) & (eV)& (eV) &\\
&&&&&\\
\hline
&&&&&\\
V1& 0.000&-3.174 & --5.818&-1.136& Al, ScN,Sc$_{2}$O$_3$ , Al$_{16}$Sc$_{8}$N$_{24}$ \\
V2&0.000& -3.172 &-5.814& -1.142& Al, AlN, Sc$_{2}$O$_3$, Al$_{16}$Sc$_{8}$N$_{24}$ \\
V3&-3.174&0.000&-3.702&-4.310&N$_{2}$, ScN, Sc$_{2}$O$_3$, Al$_{16}$Sc$_{8}$N$_{24}$\\
V4&-3.172&0.000&-3.699&-4.314&N$_{2}$, AlN,Sc$_{2}$O$_3$, Al$_{16}$Sc$_{8}$N$_{24}$ \\
&&&&&\\
\hline
\end{tabular}\\
\label{table:compositionpdwithO33}
\end{table}
%---------------------------

%---------------------------

\begin{figure}[!ht]
\centering
\includegraphics[width=0.6\linewidth]{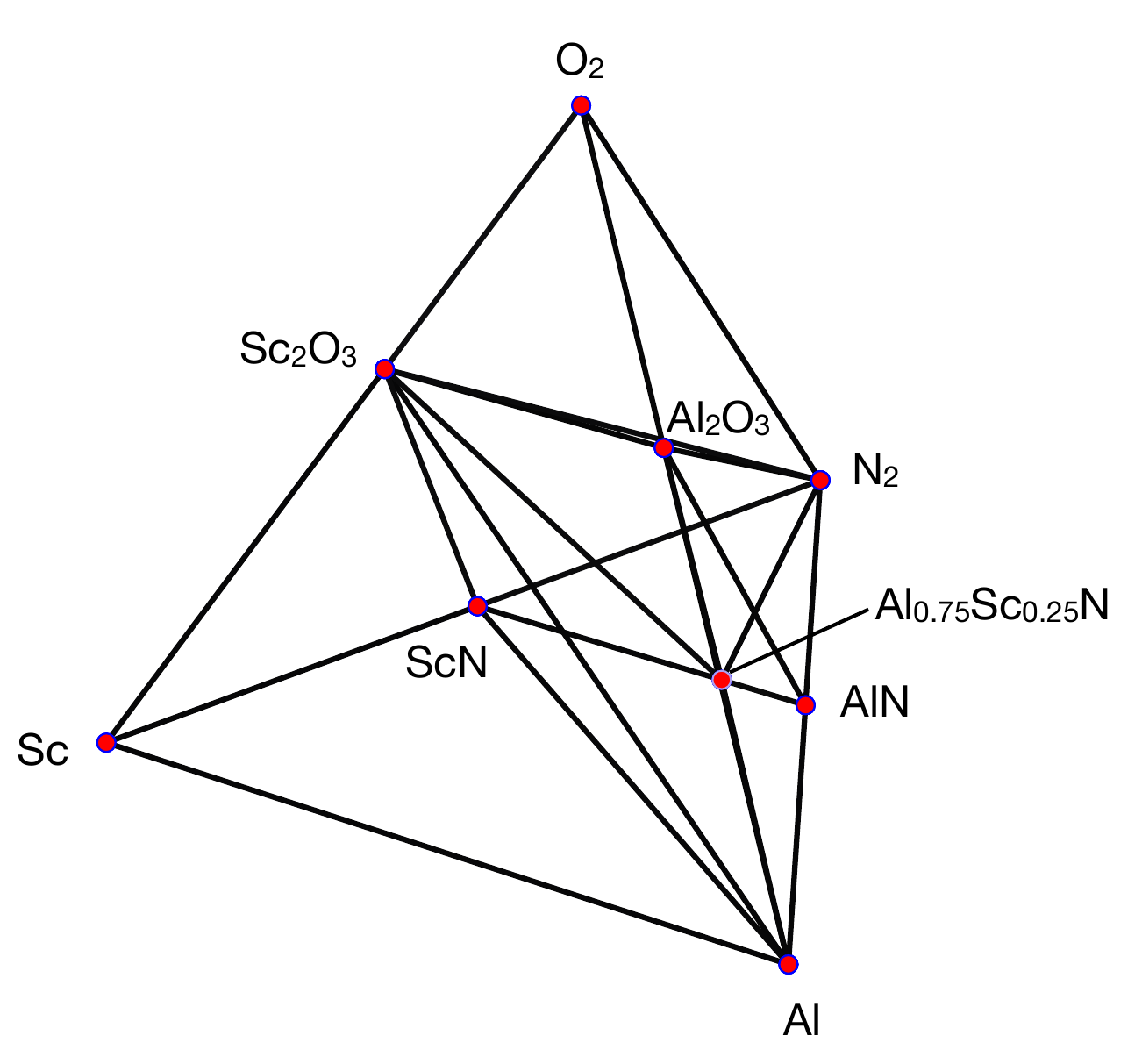}
\caption{\label{fig:compositionpdO4}
Calculated Al-Sc-N-O quaternary phase diagram showing Al$_{0.75}$Sc$_{0.25}$N as a line compound.}
\end{figure}
%---------------------------

\clearpage
%---------------------------
\section{Defect Energetics at $x$ = 0.042}

\begin{figure}[!ht]
\centering
\includegraphics[width=0.75\linewidth]{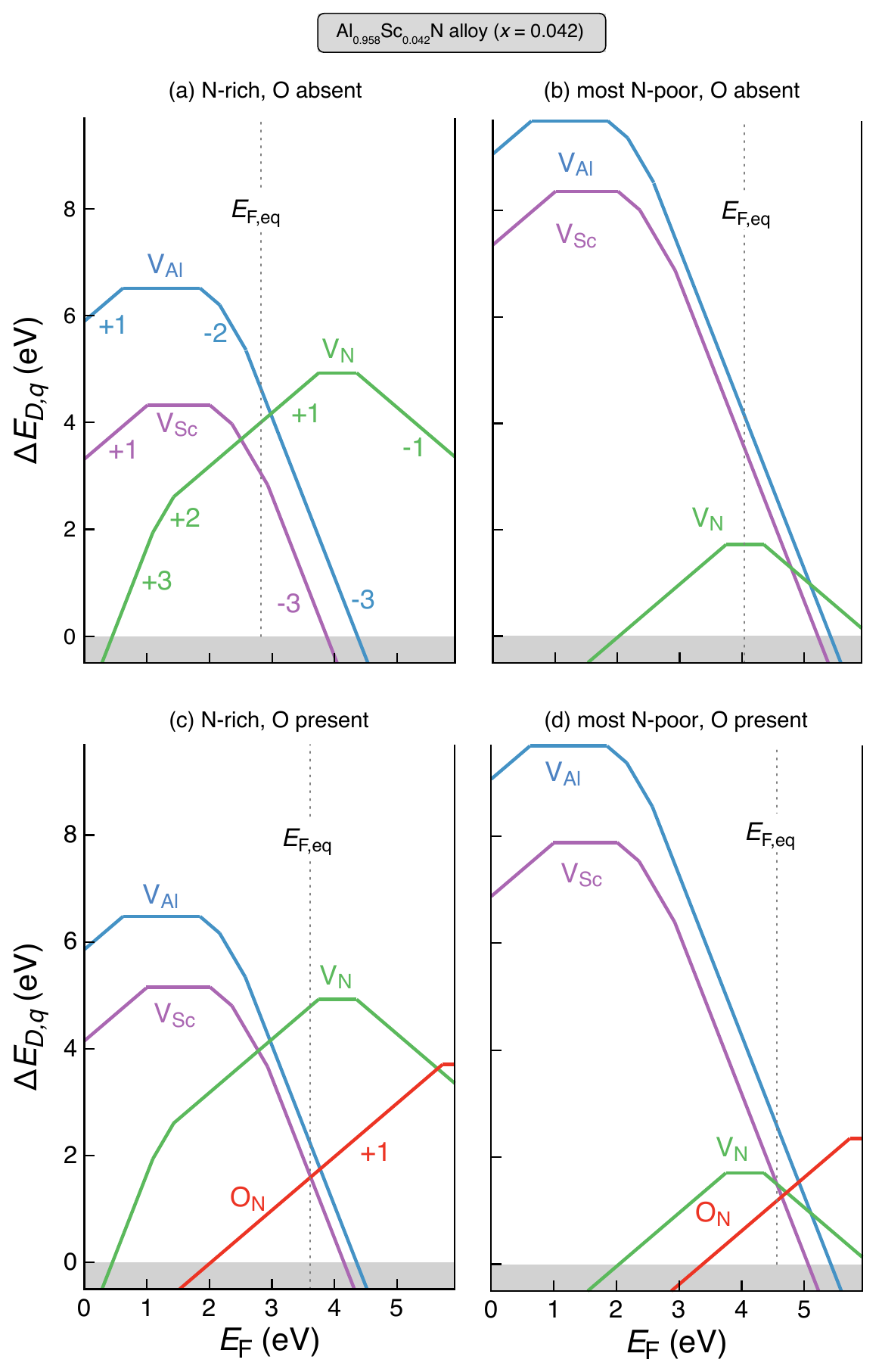}
\caption{\label{fig:SI_fig1} Calculated effective defect formation energy ($\Delta E_{D,q}$) as a function of the Fermi energy ($E_\mathrm{F}$) in Al$_{0.958}$Sc$_{0.042}$N alloys under N-rich and most N-poor growth conditions without (a, b) and with O incorporation (c, d). The equilibrium Fermi energy ($E_{\mathrm{F,eq}}$) is calculated at $T$ = 1000 K.}
\end{figure}
\clearpage

%--------------
\section{Defect Energetics at $x$ = 0.125}
\begin{figure}[!ht]
\centering
\includegraphics[width=0.75\linewidth]{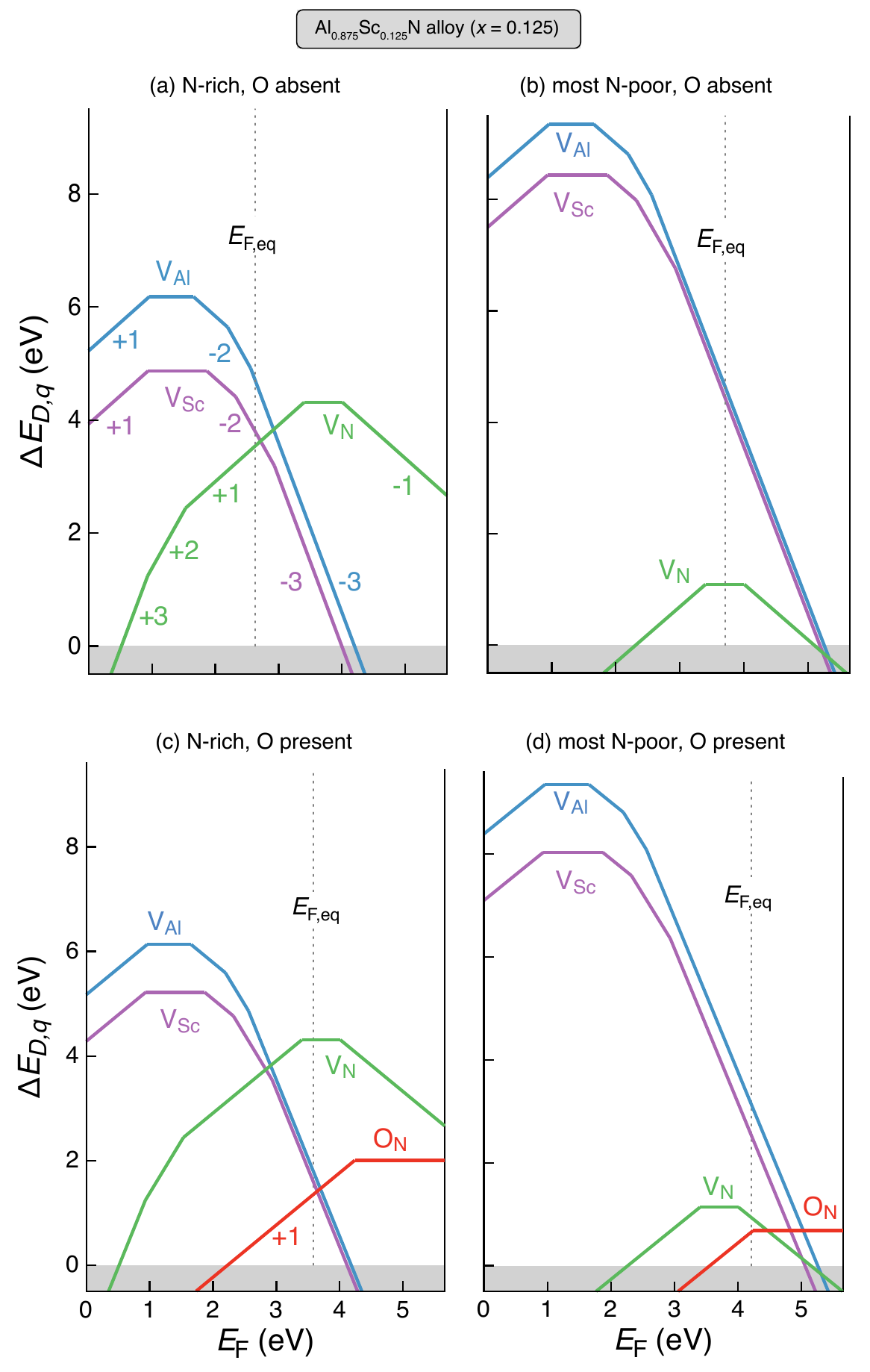}
\caption{\label{fig:SI_fig2} Calculated effective defect formation energy ($\Delta E_{D,q}$) as a function of the Fermi energy ($E_\mathrm{F}$) in Al$_{0.875}$Sc$_{0.125}$N alloys under N-rich and most N-poor growth conditions without (a, b) and with O incorporation (c, d). The equilibrium Fermi energy ($E_{\mathrm{F,eq}}$) is calculated at $T$ = 1000 K.}
\end{figure}
\clearpage

%-------------
\section{Defect Energetics at $x$ = 0.250}
\begin{figure}[!ht]
\centering
\includegraphics[width=0.75\linewidth]{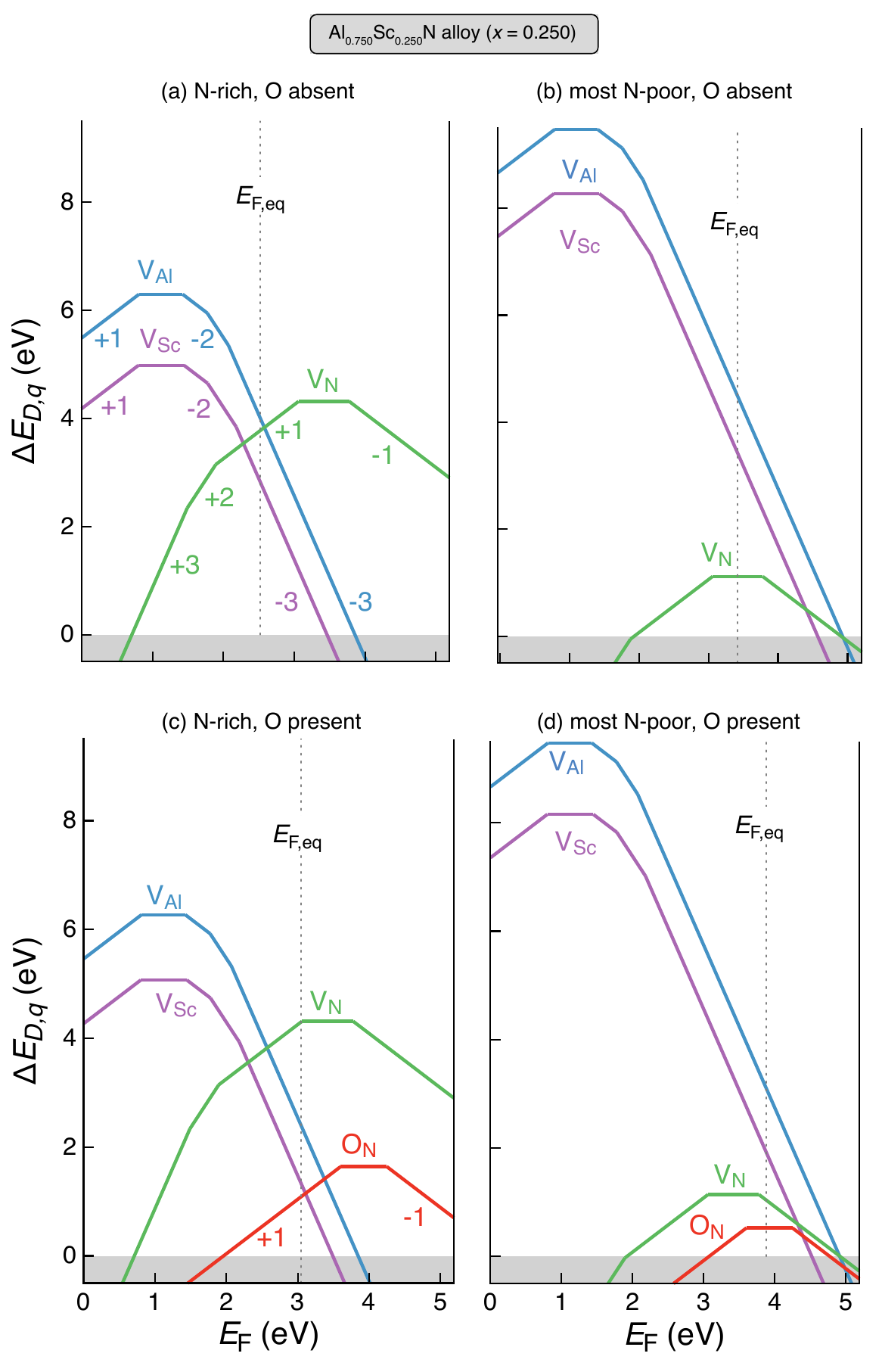}
\caption{\label{fig:SI_fig3} Calculated effective defect formation energy ($\Delta E_{D,q}$) as a function of the Fermi energy ($E_\mathrm{F}$) in Al$_{0.750}$Sc$_{0.250}$N alloys under N-rich and most N-poor growth conditions without (a, b) and with O incorporation (c, d). The equilibrium Fermi energy ($E_{\mathrm{F,eq}}$) is calculated at $T$ = 1000 K.}
\end{figure}

\clearpage
%---------------------------

\section{O$_\mathrm{N}$ DX Center in Al$_{0.67}$Sc$_{0.33}$N}
\begin{figure}[!ht]
\centering
\includegraphics[width=\linewidth]{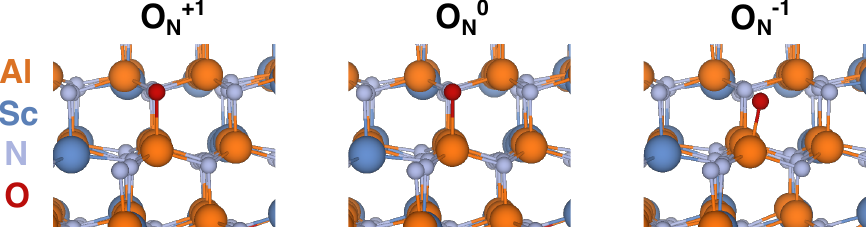}
\caption{\label{fig:SI_fig7} Local structure of substitutional O on the N site in Al$_{0.67}$Sc$_{0.33}$N alloy in defect charge states of +1, 0, and -1. O$_\mathrm{N}^{-1}$ features a large displacement of the O (red) from the N site, which is formed by capturing of two electrons by the donor O$_\mathrm{N}^{+1}$.}
\end{figure}

\clearpage
%---------------------------

%---------------------------
%---------------------------
\clearpage

%---------------------------

%---------------------------

%----REFERENCES-----
\clearpage
\bibliography{biblio}
%\bibliographystyle{rsc}
%---------------